\documentclass[3p,sort&compress]{elsarticle}

\def\ket#1{\mathinner{|{#1}\rangle}}

{\catcode`\|=\active 
  \gdef\Braket#1{\left<\mathcode`\|"8000\let|\BraVert {#1}\right>}}
\def\BraVert{\egroup\,\mid@vertical\,\bgroup}
{\catcode`\|=\active
  \gdef\set#1{\mathinner{\lbrace\,{\mathcode`\|"8000\let|\midvert #1}\,\rbrace}}
  \gdef\Set#1{\left\{\:{\mathcode`\|"8000\let|\SetVert #1}\:\right\}}}
\def\midvert{\egroup\mid\bgroup}
\def\SetVert{\egroup\;\mid@vertical\;\bgroup}

\usepackage{latexsym}
\usepackage{amssymb}
\usepackage[matrix,frame,arrow]{xy}
\usepackage{amsmath}
\newcommand{\qw}[1][-1]{\ar @{-} [0,#1]}
\newcommand{\qwx}[1][-1]{\ar @{-} [#1,0]}

\newcommand{\gate}[1]{*{\xy *+<.6em>{#1};p\save+LU;+RU **\dir{-}\restore\save+RU;+RD **\dir{-}\restore\save+RD;+LD **\dir{-}\restore\POS+LD;+LU **\dir{-}\endxy} \qw}

\newcommand{\control}{*!<0em,.025em>-=-{\bullet}}

\newcommand{\ctrl}[1]{\control \qwx[#1] \qw}
\newcommand{\rstick}[1]{*!L!<-.5em,0em>=<0em>{#1}}
\newcommand{\lstick}[1]{*!R!<.5em,0em>=<0em>{#1}}

\newcommand{\Qcircuit}[1][0em]{\xymatrix @*=<#1>}

\newcommand{\oort}{\frac{1}{\sqrt{2}}}
\newcommand{\twoVec}[2]{\begin{pmatrix}#1\\#2\end{pmatrix}}
\newcommand{\Hilb}{{\mathcal{H}}}

\def\vec#1{\mathchoice{\mbox{\boldmath$\displaystyle#1$}}
{\mbox{\boldmath$\textstyle#1$}}
{\mbox{\boldmath$\scriptstyle#1$}}
{\mbox{\boldmath$\scriptscriptstyle#1$}}}

\newtheorem{theorem}{Theorem}
\newtheorem{lemma}{Lemma}
\newdefinition{example}{Example}
\newdefinition{definition}{Definition}
\newproof{proof}{Proof}

\journal{Applied Mathematics and Computation}

\begin{document} 
	
\begin{frontmatter}

\title{De-quantisation of the Quantum Fourier Transform}
\author{Alastair A. Abbott}
\ead{aabb009@aucklanduni.ac.nz}
\address{Department of Computer Science, University of Auckland, Private Bag 92109, Auckland, New Zealand}


\begin{abstract}
The quantum Fourier transform (QFT) plays an important role in many known quantum algorithms such as Shor's algorithm for prime factorisation. In this paper we show that the QFT algorithm can, on a restricted set of input states, be de-quantised into a classical algorithm which is both more efficient and simpler than the quantum algorithm. By working directly with the algorithm instead of the circuit, we develop a simple classical version of the quantum basis-state algorithm. We formulate conditions for a separable state to remain separable after the QFT is performed, and use these conditions to extend the de-quantised algorithm to work on all such states without loss of efficiency. Our technique highlights the linearity of quantum mechanics as the fundamental feature accounting for the difference between quantum and de-quantised algorithms, and that it is this linearity which makes the QFT such a useful tool in quantum computation.
\end{abstract}


\end{frontmatter}

\section{Introduction}

The \emph{quantum Fourier transform (QFT)} plays an important role in a large number of known algorithms for quantum computers~\cite{Gruska:1999aa}. It plays a central role  in Shor's algorithm for prime factorisation~\cite{Shor:1994kx} and is often thought to be at the heart of many quantum algorithms which are faster than any known classical counterpart. However, following on from recent results relating to classical features of the QFT algorithm~\cite{Aharonov:2007aa,Browne:2007aa,Griffiths:1996aa,Yoran:2007aa}, we will argue that the QFT algorithm itself is classical in nature. 

The process of de-quantising quantum algorithms into equivalent classical algorithms is a powerful tool for investigating the nature of quantum algorithms and computation. Few general results are known about when such de-quantisations are possible and the power of quantum computation compared to classical computation. In this paper we show how the QFT algorithm can be de-quantised into a simpler, more efficient, classical algorithm when operating on a range of input states. While the de-quantised algorithms themselves are of interest, they also allow us to gain insight into the nature of the QFT. We will argue that it is the linearity inherent in the unitary quantum computational model which makes the QFT such a useful tool, rather than the nature of the QFT itself. 

In Section~\ref{sec:FT_bg} of this paper we overview the basic QFT theory and present the QFT algorithm in a compact form which allows us to move away from the restrictions imposed by the circuit layout. In Section~\ref{sec:initDe-quant} we overview the de-quantisation procedure and de-quantise the QFT algorithm acting on a basis-state input. In Section~\ref{sec:prod_deq} we explore the entangling power of the QFT and determine conditions for when a separable input state remains unentangled by the QFT, before presenting a de-quantised algorithm that works on such product-state inputs. In Section~\ref{sec:discuss} we discuss why de-quantisation of the QFT is possible and note some common misunderstandings about the QFT which contribute to this.

\section{Discrete and Quantum Fourier Transforms}\label{sec:FT_bg}

The \emph{discrete Fourier transform (DFT)} on which the QFT is based is a transformation on a $q$-dimensional complex vector $\chi = (f(0),f(1),\dots,f(q-1))$ into its Fourier representation $\hat{\chi}=(\hat{f}(0),\hat{f}(1),\dots,\hat{f}(q-1))$~\cite{Gruska:1999aa}:
\begin{equation}
	\label{eqn:DFT_defn}
	\hat{f}(c)=\frac{1}{\sqrt{q}}\sum_{a=0}^{q-1}e^{2\pi iac/q}f(a) ,
\end{equation}
for $c\in \{0,1,\dots,q-1\}$. The QFT is similarly defined so that the transformation acts on a state vector in $q$-dimensional Hilbert space, $\mathcal{H}_q$. In quantum computation we work with a state vector defining a register comprising of $n$ two-state qubits, so we will only consider the case that $q=2^n$ from this point onwards. We will use the convention that $n$ is the number of qubits while $N=2^n$ is the dimension of Hilbert space the $n$ qubits are in. This means that the QFT, denoted $F_q$, acts on the $N$ amplitudes of a particular $n$-qubit state, i.e. 
\begin{equation}
	\label{eqn:QFT_defn}
	\sum_{a=0}^{N-1}f(a)\ket{a} \xrightarrow{F_{N}} \sum_{c=0}^{N-1}\hat{f}(c)\ket{c} .
\end{equation}
The QFT hence transforms a state so as to perform a DFT on its state vector.

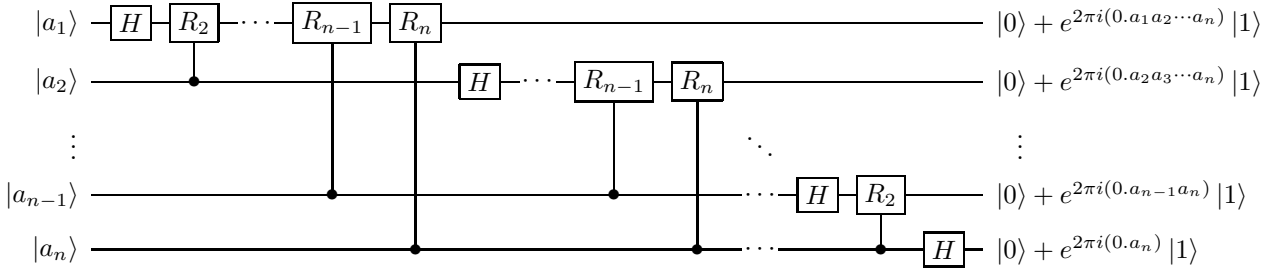
\begin{figure}[t]
	\scalebox{1}{
	\qquad
	\Qcircuit @C=.7em @R=.7em {
			&\lstick{\ket{a_1}} & \gate{H} & \gate{R_2} & \qw & \cdots & & \gate{R_{n-1}} & \gate{R_n} & \qw & \qw & \qw & \qw & \qw & \qw & \qw & \qw & \qw & \qw & \qw & \qw & \rstick{\ket{0} + e^{2\pi i(0.a_1 a_2 \cdots a_n)}\ket{1}} \qw\\
			&\lstick{\ket{a_2}} & \qw & \ctrl{-1} & \qw & \qw & \qw & \qw & \qw & \gate{H} & \qw & \cdots & & \gate{R_{n-1}} & \gate{R_n} & \qw & \qw & \qw & \qw & \qw & \qw & \rstick{\ket{0} + e^{2\pi i(0.a_2 a_3 \cdots a_n)}\ket{1}} \qw\\
			&&&&&&&&&&&&&&&&&&&&&\\
			\vdots&&&&&&&&&&&&&&&&\ddots&&&&&&&\vdots\\
			&&&&&&&&&&&&&&&&&&&&&\\
			&\lstick{\ket{a_{n-1}}} & \qw & \qw & \qw & \qw & \qw & \ctrl{-5} & \qw & \qw & \qw & \qw & \qw & \ctrl{-4} & \qw & \qw & \cdots  &  & \gate{H} & \gate{R_2} & \qw & \rstick{\ket{0} + e^{2\pi i(0.a_{n-1} a_n)}\ket{1}} \qw\\
			&\lstick{\ket{a_n}} & \qw & \qw & \qw & \qw & \qw & \qw & \ctrl{-6} & \qw & \qw & \qw & \qw & \qw & \ctrl{-5} & \qw & \cdots &  & \qw & \ctrl{-1} & \gate{H} & \rstick{\ket{0} + e^{2\pi i(0.a_n)}\ket{1}} \qw\\
		}
	}
	\caption[The quantum circuit computing the quantum Fourier transform.]{The standard quantum circuit for the QFT. The output normalisation factors of $1/\sqrt{2}$ and swap gates to reverse qubit order are omitted.}
	\label{fig:qft:qftCircuit}
\end{figure}

As a result of the linearity of quantum mechanics, in order to compute the QFT we only need to design an algorithm to transform a single component of the state vector. This is because an arbitrary state $\ket{\psi_N} = \sum_{a=0}^{N-1}f(a)\ket{a}$ transforms as:
$$F_{N}\ket{\psi_N} = \sum_{a=0}^{N-1}f(a)F_{N}\ket{a} =  \frac{1}{\sqrt{N}}\sum_{a=0}^{N-1}\sum_{c=0}^{N-1}e^{2\pi iac/N}f(a)\ket{c} = \sum_{c=0}^{N-1}\hat{f}(c)\ket{c} .$$
Hence we arrive at the standard definition of the QFT as the mapping~\cite{Cleve:1997aa}
\begin{equation}
	\label{eqn:QFT_SingleStateAction}
	\ket{a} \xrightarrow{F_{N}} \frac{1}{\sqrt{N}}\sum_{c=0}^{N-1}e^{2\pi iac/N}\ket{c} ,
\end{equation}
with $a \in \{0,1,\dots,N-1\}$. 
Following the standard procedure~\cite{Cleve:1997aa}, we proceed to decompose \eqref{eqn:QFT_SingleStateAction} into a separable form.
Keeping in mind that we are dealing with registers composed of qubits, we can decompose $a$ (and similarly $c$) into its binary representation so that $a=2^{n-1}a_1 + 2^{n-2}a_2 + \cdots + 2^1 a_{n-1} + 2^0 a_n$ and $\ket{a} = \ket{a_1 a_2 \cdots a_n}$. By denoting $a=a_1 a_2 \cdots a_n$ and $a/2^n = 0.a_1 a_2 \cdots a_n$ we observe that
\begin{align}
	\label{eqn:QFT:qftExp}
	e^{2\pi iac/2^n} &= e^{2\pi ia(2^{n-1}c_1 + 2^{n-2}c_2 + \cdots + 2^0 c_n)/2^n}\notag\\
	&= e^{2\pi i (a_1a_2\cdots a_n) c_1/2^{1} }e^{2\pi i (a_1a_2\cdots a_n)c_2/2^{2} }\cdots e^{2\pi i (a_1a_2\cdots a_n)c_n/2^n }\notag\\
	&= e^{2\pi i(a_1\cdots a_{n-1}.a_n)c_1}e^{2\pi i(a_1 \cdots a_{n-2}.a_{n-1}a_n)c_2}\cdots e^{2\pi i(0.a_1 a_2\cdots a_n)c_n} .
\end{align}
Noting that for any decimal $x.y$ we have $e^{2\pi i (x.y)} = (e^{2\pi i})^x e^{2\pi i(0.y)} = e^{2\pi i(0.y)}$, we see that only the fractional part of $(a_1 \cdots a_{n-j}.a_{n-j+1}\cdots a_n)c_j$ is of any significance in the exponent of \eqref{eqn:QFT:qftExp}.\footnote{This technique of removing factors of $(e^{2\pi i})^k$ for $k\in \mathbb{N}$ will be commonly used throughout this paper to reduce formulae.} Hence, we find
\begin{equation*}
	e^{2\pi iac/2^n}\ket{c_1\cdots c_n} =  e^{2\pi i(0.a_n)c_1}\ket{c_1}\cdots e^{2\pi i(0.a_1 a_2\cdots a_n)c_n}\ket{c_n} .
\end{equation*}
Using this decomposition we can write \eqref{eqn:QFT_SingleStateAction} as a product state of individual qubits,
\begin{align}
	\label{eqn:QFT_separated}
	\sum_{c=0}^{N-1}e^{2\pi iac/2^n}\ket{c} = (\ket{0} + e^{2\pi i(0.a_n)}\ket{1})\cdots
	(\ket{0} + e^{2\pi i(0.a_1\cdots a_n)}\ket{1}) .
\end{align}
The quantum algorithm to implement the QFT follows directly from this factorisation. The circuit for the algorithm is shown in Figure~\ref{fig:qft:qftCircuit}. The algorithm can be written explicitly as follows~\cite{Cleve:1997aa}:

\vspace{10pt}\noindent
{\bfseries Quantum Fourier Transform}\\
{\bfseries Input:} The state $\ket{a} = \ket{a_1}\ket{ a_2} \cdots \ket{a_n}$.\\
{\bfseries Output:} The transformed state $\frac{1}{\sqrt{N}}(\ket{0} + e^{2\pi i(0.a_n)}\ket{1})\cdots (\ket{0} + e^{2\pi i(0.a_1 \cdots a_n)}\ket{1})$.
\begin{enumerate}
	\item For $j=1$ to $n$, transform qubit $\ket{a_j}$ as follows:
	\item \quad $\ket{a_j} \xrightarrow{H} \oort (\ket{0} + e^{2\pi i(0.a_j)}\ket{1})$.
	\item \quad For $k=j+1$ to $n$:
	\item \quad \quad $\oort (\ket{0} + e^{2\pi i(0.a_j\cdots a_{k-1})}\ket{1}) \xrightarrow {R_k} \oort (\ket{0} + e^{2\pi i(0.a_j\cdots a_{k-1}a_k)}\ket{1})$ where $R_k$\\
	\phantom{.}\quad \quad is the unitary $k$-controlled phase shift:
	$$R_k = 
	\begin{pmatrix}
		1 & 0 & 0 & 0\\
		0 & 1 & 0 & 0\\
		0 & 0 & 1 & 0\\
		0 & 0 & 0 & e^{2\pi i/2^k}\\
	\end{pmatrix} .$$
	\item \quad End For.
	\item \quad Reverse the order of the qubits.
	\item End For.
\end{enumerate}
Clearly this produces the state \eqref{eqn:QFT_separated} and requires $O(n^2)$ unitary $R_k$ and $H$ gates to run.

There are a few important notes about the QFT which should be made. While both the DFT and the QFT act on vectors in a complex vector space, the DFT acts on an abstract, mathematical vector, whereas the QFT acts on a physical state which we mathematically represent by a vector in $\Hilb_{N}$. The subtle difference here is that with the classical DFT, we can read the values of all $2^n$ Fourier coefficients $\hat{f}(c)$ by simple inspection of the transformed vector. With the QFT, the resulting state \eqref{eqn:QFT_defn} embeds all $2^n$ coefficients as amplitudes for the $2^n$ states of  an $n$-qubit system. However, the collapse of the superposition upon measurement means that it is impossible to measure the amplitudes of a quantum state without an ensemble of such states to make a statistical approximation of the amplitudes from~\cite{Jozsa:1997aa}. Hence, the quantum state \eqref{eqn:QFT_defn} contains all the information of the classically transformed vector, but it is inaccessible to measurement. The main use of the QFT is then as a tool to extract information embedded in the relative amplitudes of states as opposed to determining the coefficients themselves. 

Another result of this is that the efficiency of the QFT ($O(n^2)$ as opposed to the DFT which is $O(n2^n)$) is in some sense due to the ability to perform the transformation and utilise the information in the phases without measuring the state. Evidently, any algorithm requiring measurement needs exponential time (there are $2^n$ coefficients to measure), so even if quantum mechanics would allow us to measure the Fourier coefficients in state \eqref{eqn:QFT_defn}, doing so would take $O(n2^n)$ time: $2^n$ coefficients, $n$ qubits each. Making use of this embedded information while avoiding measurement is certainly an important part of the fine art of developing algorithms in quantum computing.

\section{Initial De-quantisation Investigation} \label{sec:initDe-quant}

Having presented the QFT, there are some issues to be brought to light. The decomposition of the transformed state \eqref{eqn:QFT_SingleStateAction} (shown in \eqref{eqn:QFT_separated}) is evidently not entangled, and the separability of the state would lead us to believe that the QFT algorithm producing it could be simulated efficiently in a classical manner~\cite{Abbott:2009aa,Jozsa:2003aa}, and there are certainly results towards this.

It was realised shortly after the discovery of Shor's algorithm that the QFT could be computed in a semiclassical manner~\cite{Griffiths:1996aa}. By using classical signals resulting from quantum measurements, one can perform the QFT on a state using classical logic and one-qubit gates (instead of the usual two-qubit controlled-phase-shifts). This method gives the same resulting probability distribution as the quantum algorithm, but destroys the state's superposition as it relies on irreversible measurements. As a result, this is only useful in an algorithm in which the QFT directly precedes measurement. Shor's algorithm happens to be of exactly this nature, but this is only an initial step towards true classical simulation. 

Much more recently, classical simulations of the QFT have been studied from the viewpoint of simulating the circuit in Figure~\ref{fig:qft:qftCircuit} by exploiting the bubble-width of the quantum circuit~\cite{Aharonov:2007aa} and the tensor contraction model~\cite{Yoran:2007aa}. The bubble-width approach uses a slightly modified version of the QFT circuit which is of logarithmic bubble-width and simulates this circuit. The tensor-contraction model also focuses on the circuit topology, but relies on associating a tensor with each vertex in the circuit, then cleverly contracting the tensors into a single rank-one tensor. Both these methods work on separable input states, but are sampling-based forms of de-quantisation~\cite{Abbott:2010aa} in the sense that a final measurement is assumed and an output is classically sampled from the correct (calculated) probability distribution. This makes these de-quantisations less general than might be desired and difficult to apply when the QFT is used, as it often is, as a part of a larger quantum algorithm. This is because in these cases measurement cannot be assumed after the QFT, and the de-quantisation must be cleverly and non-trivially composed with a de-quantisation of the rest of the algorithm to be applied.


 Working with the circuit topology, while beneficial for some purposes, also seems to overcomplicate matters and restrict generalisation when it comes to classical simulation. We will explore simulations of the QFT in a different light, more along the lines of the de-quantisation explored previously by Abbott~\cite{Abbott:2009aa} and Calude~\cite{Calude:2007aa} which aim to provide stronger (not sampling-based) de-quantisations when possible.

\subsection{De-quantisation Overview}

The idea behind this de-quantisation procedure is that qubits which are separable exhibit only superposition and interference. These properties are the result not of non-classical features of the qubits, but rather of the two-dimensionality of the qubits. By using classical, deterministic two-dimensional bits instead of qubits, the same behaviour can be exhibited without the difficulties imposed by measurement and probabilities. Not all algorithms fit within this paradigm, but there are many which can be tackled with this approach.  Algorithms which use measurement as a fundamental part of their procedure are examples of those which are not so well suited, and sampling-based techniques are more suitable in these situations. Finding when these stronger de-quantisations are possible also gives insight into the power of particular quantum algorithms~\cite{Abbott:2010aa}, as this reflects to some degree the amount by which the algorithm utilises the possible advantages of quantum mechanics. In cases where entanglement is bounded \cite{Jozsa:2003aa}, we can use this de-quantisation procedure to produce classical algorithms which are as efficient as their quantum counterparts. This procedure was explicitly examined further~\cite{Abbott:2009aa,Calude:2007aa} when applied to the Deutsch-Jozsa problem \cite{Deutsch:1985aa,Deutsch:1992aa}, where complex numbers were used as classical two-dimensional bits. In this paper we will apply this de-quantisation procedure to the QFT, but because the amplitudes we need to represent in the QFT algorithm are complex-valued, we cannot use complex numbers as our two-dimensional bits. There is no problem though with simply using two-valued vectors as our classical bits, so we will employ this procedure.

\subsection{Basis-state De-quantisation}

The de-quantisation for a basis-state needs only to simulate the transformation defined in \eqref{eqn:QFT_SingleStateAction}. As a result of the decomposition in \eqref{eqn:QFT_separated}, the effect of the QFT on the $j$th qubit is easily seen to be
\begin{equation}
	\label{eqn:QFT_qubit_def}
	\ket{a_j} \xrightarrow{F_{2^n}} \oort (\ket{0} + e^{2\pi i(0.a_{n-j+1}\cdots a_n)}\ket{1}) .
\end{equation}
The difficulty in implementing this in a quantum computer is that the phase of a qubit needs to be altered depending on the values of the other qubits without altering them
and the circuit of controlled-phase-shifts is required to implement this. The information is spread over the input qubits and must be obtained without measurement. In the classical case there are no such restrictions on measurement, so de-quantisation should only require directly implementing \eqref{eqn:QFT_qubit_def}. However, evaluating the complex phase for each of the $n$ qubits takes $O(n)$ time, leading to a $O(n^2)$ procedure. This can be reduced to $O(n)$ by calculating each phase dependent on the previous one. To do so, let $\omega_j$ be the $j$th phase factor and note the following:
\begin{align}
	\omega_j &= e^{2\pi i (0.a_{n-j+1}\cdots a_n)}\notag\\
			 &= e^{2\pi i (0.a_{n-j+1})}e^{2\pi i (0.a_{n-j+2}\cdots a_n)/2}\notag\\
			 &= (-1)^{a_{n-j+1}}\sqrt{\omega_{j-1}} \notag,
\end{align}
and $$\omega_1 = e^{2\pi i(0.a_n)} = (-1)^{a_n} ,$$
where by the square-root we mean the principal root. The square-root of a complex number such as $\omega_j$ can be calculated independently of $n$. Specifically, if we have $s+ti = \sqrt{b + di}$ with the further requirement that for a root of unity $\sqrt{b^2+d^2} = 1$, then~\cite{Barnard:1936aa}:
\begin{align*}
	s &= \oort \sqrt{1+b}, \quad t = \frac{\mbox{sgn}(d)}{\sqrt{2}}\sqrt{1-b},
\end{align*}
where sgn$(d) = d/|d|$ is the sign of $d$.
The efficient de-quantised algorithm is then the following:

\vspace{10pt}\noindent
{\bfseries Basis-state De-quantised QFT}\\
{\bfseries Input:} The binary string $a=a_1 a_2 \dots a_n$.\\
{\bfseries Output:} The $n$ transformed two-component complex vectors $\vec{b_1} \vec{b_2} \dots \vec{b_n}$.
\begin{enumerate}
	\item Let $\omega = 1$
	\item For $j=1$ to $n$:
	\item \quad Set $\omega = (-1)^{a_{n-j+1}}\sqrt{\omega}$
	\item \quad Set $\vec{b_j} = \oort \times \twoVec{1}{\omega}$
	\item End For
\end{enumerate}
This algorithm produces vectors mathematically identical to the state-vectors in \eqref{eqn:QFT_SingleStateAction} and \eqref{eqn:QFT_separated} produced by the QFT, but is computed classically in $O(n)$ time -- more efficient than the quantum solution and simpler too. This is primarily because the quantum circuit is constructed subject to the requirement of computing the QFT without any intermediate measurements. As a result, the quantum algorithm corresponding to the circuit must conform to this too, making it more complex than an equivalent classical algorithm need be.

A classical algorithm has the further advantage over the quantum algorithm acting on a basis-state that measurement of the resulting state can be performed at will, and any required information is easily accessible. In the quantum algorithm only a single state can be measured, and no information about the amplitudes (and thus the Fourier coefficients) can be determined from a single QFT application. While this classical algorithm is no faster than the well known fast Fourier transform (FFT) for calculating all the coefficients, it may be advantageous if only some coefficients are required.

The ability to de-quantise the QFT acting on a basis state is not particularly surprising. This is equivalent to the classical DFT acting on a vector with only one non-zero component, producing a fairly trivial and easily computed output. However, this highlights a little more deeply some common misconceptions about the QFT. Because of the linear, unitary evolution of quantum mechanics, the action of the QFT on a basis state shown in \eqref{eqn:QFT_SingleStateAction} is often taken as the definition of the QFT. While this suffices as the definition for the purposes of the quantum algorithm, it is important not to forget that the actual definition of the QFT is that given in \eqref{eqn:QFT_defn}. When considering classical simulations of the QFT this is even more important, as the action of the QFT on a basis state and the corresponding circuit no longer immediately allow us to compute the complete QFT; indeed it would take $2^n$ iterations of a classical algorithm simulating the basis state behaviour to compute the complete QFT. 

\section{Product-state De-quantisation}\label{sec:prod_deq}

Here we consider the possibility of extending the de-quantisation to work on a wider range of input states, resulting in a less trivial de-quantisation. If the input state is entangled then it is clear that the de-quantisation is not easily extended, as the method used for the basis-state algorithm relied on the separability of the input. In such a situation, any de-quantisation attempt would need to involve a different method and work directly from the QFT definition, \eqref{eqn:QFT_defn}.

It is not immediately clear that the basis-state de-quantisation, which is based on \eqref{eqn:QFT_SingleStateAction}, could not be extended to work on arbitrary separable input states. This idea is strengthened by the fact that we used the single-qubit formula \eqref{eqn:QFT_qubit_def} to perform the basis-state de-quantisation. However, this implicitly relies on the other qubits in the input state having a definite value, but in the general separable input case this is not necessarily the case. Indeed, the QFT is readily seen to entangle separable input states, e.g.: \begin{equation*}
\label{eqn:QFTentEx}
\ket{\phi} = \oort\ket{0}\left(\ket{0} + \ket{1}\right) \xrightarrow{F_4} \oort\left(\ket{00} + \frac{1+i}{2}\ket{01} + \frac{1-i}{2}\ket{11}\right).
\end{equation*}
A de-quantisation for arbitrary separable input states is thus not possible in the same way as it was for basis states. However, we will investigate the entangling power of the QFT in order to determine the set of states which are not entangled by the QFT, and present a de-quantised algorithm which works for such states.

\subsection{General Separability Conditions}

As in the entanglement investigation of the Deutsch-Jozsa problem~\cite{Abbott:2009aa}, we will make use of the separability conditions for a qubit state presented in \cite{Jorrand:2003aa}, although unlike the Deutsch-Jozsa problem our situation permits the possibility of states with zero-valued amplitudes, complicating the conditions somewhat. The key definitions and theorems we require to determine the separability of a state will be briefly presented, while \cite{Jorrand:2003aa} should be consulted for proofs and discussion.

\begin{definition}
	The \emph{amplitude abstraction function} $\mathcal{A} : \Hilb_{N} \to \{0,1\}^N$ is a function which, when applied to a state $\ket{\psi_N} = \sum_{i=0}^{N-1}c_i\ket{i}$, yields a bit string $x=x_0 x_1 \dots x_{N-1}$ such that for $0\le i \le N-1$, $x_i=0$ if $c_i=0$ and $x_i=1$ otherwise.
\end{definition}
\begin{definition}\label{def:Bn}
	The set $\mathcal{B}_N \subset \{0,1\}^N$ of \emph{well-formed bit strings} of length $N=2^n$ is defined recursively as
	\begin{align*}
		\mathcal{B}_2 &= \{01,10,11\}, \quad
		\mathcal{B}_{2N} = \{0^N x, x0^N, xx \mid x\in \mathcal{B}_N \} .
	\end{align*}
\end{definition}
\begin{definition}
	The set of \emph{well-formed states} is the set $$\mathcal{V}_N = \{\ket{\psi_N} \in \Hilb_N \mid \mathcal{A}\left(\ket{\psi_N}\right) \in \mathcal{B}_N\} .$$
\end{definition}
Intuitively, a state is well-formed if the zero-valued amplitudes are distributed such that it is a candidate to be separable; if a state is not well-formed it is guaranteed to be entangled. In order to determine if a well-formed state is separable, we require two further definitions.
\begin{definition}
	For each set of well-formed states $\mathcal{V}_N$, there exists a family of \emph{zero deletion functions} $\{\mathcal{D}_K : \mathcal{V}_N \to \Hilb_K \mid K=2^k, 1 \le k \le n \}$, such that for a well-formed state $\ket{\psi_N} = \sum_{i=0}^{N-1}c_i\ket{i} \in \mathcal{V}_N$, $\mathcal{D}_K(\ket{\psi_N}) = \ket{\psi'_K}=\sum_{j=0}^{K-1}c'_j\ket{j}$, $\mathcal{A}\left(\ket{\psi'_K}\right)=1^K$, and $c'_j$ is the $j$th non-zero amplitude of $\ket{\psi_N}$.
\end{definition}
\begin{definition}
	\label{def:PPI}
	A state $\ket{\psi_N} = \sum_{i=0}^{N-1}c_i \ket{i}$ is \emph{pair product invariant} if and only if for all $j \in \{2,\dots,n\}$ and all $m \in \{0,\dots,J/2-1\}$  $c_m c_{J-m-1} = d_j$, where each $d_j$  is a constant and $J=2^j$.
\end{definition}
As a concrete example to help understand pair product invariance, consider the cases of $n=2$ and $n=3$. For $n=2$, $\ket{\psi_4}=\sum_{i=0}^3 c_i\ket{i}$ is pair product invariant if the well known condition $c_0 c_3 = c_1 c_2$ holds. For $n=3$, $\ket{\psi_8}=\sum_{i=0}^7 c_i\ket{i}$, we require this same condition, $c_0 c_3 = c_1 c_2$, as well as the further condition that $c_0 c_7 = c_1 c_6 = c_2 c_5 = c_3 c_4$, to hold.

The following theorem from \cite{Jorrand:2003aa} can be used to determine if an arbitrary $n$-qubit state is separable or not by checking the non-zero amplitudes of the state vector are pair product invariant. 
\begin{theorem}
	\label{thm:genSepCond}
	Let $\ket{\psi_N}$ be an $n$-qubit state for which the bit string $\mathcal{A}\left(\ket{\psi_N}\right)$ contains $K$ ones. Then $\ket{\psi_N}$ is separable if and only if $\ket{\psi_N}\in \mathcal{V}_N$ and $\mathcal{D}_K\left(\ket{\psi_N}\right)$ is pair product invariant.
\end{theorem}

In order to help grasp these concepts which will be critical in the rest of the paper, we present two examples.
\begin{example}
	Consider the state $$\ket{\psi_8}=\left(\frac{2i}{\sqrt{35}}\raisebox{0.9mm}{,}\frac{-4}{\sqrt{105}}\raisebox{0.9mm}{,}\frac{1}{\sqrt{35}}\raisebox{0.9mm}{,}\frac{2i}{\sqrt{105}}\raisebox{0.9mm}{,}-2\sqrt{\frac{2}{35}}\raisebox{0.9mm}{,}-4i\sqrt{\frac{2}{105}}\raisebox{0.9mm}{,}i\sqrt{\frac{2}{35}}\raisebox{0.9mm}{,}-2\sqrt{\frac{2}{105}}\right)^T\raisebox{0.9mm}{.}$$
	The state has no zero-valued amplitudes, so to check if it is separable we must simply check that it is pair product invariant. It is easily seen that we have
	$$\frac{2i}{\sqrt{35}}\times -2\sqrt{\frac{2}{105}} = \frac{-4}{\sqrt{105}}\times i\sqrt{\frac{2}{35}} = \frac{1}{\sqrt{35}}\times -4i\sqrt{\frac{2}{105}} = \frac{2i}{\sqrt{105}}\times -2\sqrt{\frac{2}{35}} = \frac{-4i}{35}\sqrt{2}{3},$$
	and also
	$$\frac{2i}{\sqrt{35}}\times \frac{2i}{\sqrt{105}} = \frac{-4}{\sqrt{105}}\times \frac{1}{\sqrt{35}} = \frac{-4}{35\sqrt{3}}\raisebox{0.9mm}{,}$$
	so $\ket{\psi_8}$ is pair product invariant and thus separable. This procedure is powerful as it is by no means clear a priori that the state is separable; indeed, small modifications to the amplitudes (e.g. swapping the -4 and 2 in the first two amplitudes) yield almost identical states, but which are not separable.
\end{example}
\begin{example}
	Consider the state
	$$\ket{\phi_8} = \left( \frac{1-i}{2\sqrt{2}}\raisebox{0.9mm}{,0,} \frac{1}{2}\raisebox{0.9mm}{,0,} \frac{i}{2}\raisebox{0.9mm}{,0,} \frac{i-1}{2\sqrt{2}}\raisebox{0.9mm}{,0} \right)^T\raisebox{0.9mm}{.}$$
	We have $\mathcal{A}(\ket{\phi_8})=10101010\in \mathcal{B}_8$, so $\ket{\phi_8}\in\mathcal{V}_8$. Since the state is well formed, the zero-deleted state is 
	$$\mathcal{D}_4(\ket{\phi_8})=\ket{\phi_8} = \left( \frac{1-i}{2\sqrt{2}}\raisebox{0.9mm}{,} \frac{1}{2}\raisebox{0.9mm}{,} \frac{i}{2}\raisebox{0.9mm}{,} \frac{i-1}{2\sqrt{2}}\raisebox{0.9mm}{,} \right)^T\raisebox{0.9mm}{.}$$
	A quick check verifies that
	$$\frac{1-i}{2\sqrt{2}}\times \frac{i-1}{2\sqrt{2}} = \frac{1}{2}\times \frac{i}{2} = \frac{i}{4}\raisebox{0.9mm}{,}$$
	and $\mathcal{D}_4(\ket{\phi_8})$ is pair product invariant and thus $\ket{\phi_8}$ is separable.
\end{example}

\subsection{QFT Separability Conditions}\label{sec:prodDeq:qftSep}

We wish to consider the case that a separable $n$-qubit input state remains separable after the QFT has been applied to it. In order to do so, first let us consider the action of the QFT on the separable input state
\begin{align*}
	\ket{\psi_N} &= \twoVec{\alpha_1}{\beta_1}\otimes \twoVec{\alpha_2}{\beta_2} \otimes \cdots \otimes \twoVec{\alpha_{n}}{\beta_{n}}
	= \left(f(0),f(1),\dots,f(N-1)\right)^\mathrm{T} .
\end{align*}
Note that each $f(c)$ can be written as a product of amplitudes as $f(c)=a_1 a_2 \dots a_n$, where each $a_i \in \{\alpha_i,\beta_i\}$. We will use the notation $f_j(c)$ to mean $a_j a_{j+1} \dots a_n$, and thus $f(c) = f_1(c) = a_1 f_2(c)$ etc. Because of the structure of the tensor product, for $0 < j < n$ and $c < 2^{n-j}$, $f_j(c) = \alpha_j f_{j+1}(c)$ and $f_j(2^{n-j} + c) = \beta_j f_{j+1}(c)$. The amplitudes of the transformed state $\ket{\hat{\psi}_N} = (\hat{f}(0), \hat{f}(1), \dots, \hat{f}(N-1))^\mathrm{T}$ are given by \eqref{eqn:DFT_defn}, which can, for a separable input, be rewritten in the more useful form
\begin{align}
	\hat{f}(c) &= \frac{1}{\sqrt{N}}\sum_{a=0}^{N-1}e^{2\pi i a c/N} f_1(a)\notag\\
	&= \frac{1}{\sqrt{N}}\alpha_1 \sum_{a=0}^{N/2-1}e^{2\pi i a c/N}f_2(a) + \beta_1 \sum_{a=0}^{N/2-1}e^{2\pi i (N/2+a) c/N}f_2(a)\notag\\
	&= \frac{1}{\sqrt{N}}(\alpha_1 + e^{\pi i c}\beta_1)\sum_{a=0}^{N/2-1}e^{2\pi i a c/N}f_2(a)\notag\\
	&= \frac{1}{\sqrt{N}}(\alpha_1 + e^{\pi i c}\beta_1)(\alpha_2 + e^{\pi i c/2}\beta_2)\cdots (\alpha_n+e^{\pi i c/2^{n-1}}\beta_n)\notag\\
	&= \frac{1}{\sqrt{N}}\prod_{j=1}^n(\alpha_j + e^{\pi i c/2^{j-1}}\beta_j)\label{eqn:factorisedQFTOuputS}.
\end{align}

This factorised form of the transformed Fourier coefficients allows us to determine conditions for when the transformed state is well-formed by giving restrictions on the distribution of zeros amongst the amplitudes, and is a significant step towards determining if a state is separable, and thus de-quantisable.
More specifically, we note that the products of the first $k$ factors in \eqref{eqn:factorisedQFTOuputS} are equal for $c=m2^j+d$ where $d<2^j$, $0\le m \le 2^{n-j}-1$. This introduces certain symmetries between amplitudes which we will exploit in the proofs which follow.
For example, since we have $\hat{f}(c) = 0$ if and only if one of the factors in \eqref{eqn:factorisedQFTOuputS} is zero, if $\hat{f}(c)=0$ for some $c<2^j$ then we must also have $\hat{f}(m2^j+c)=0$ for $0\le m \le 2^{n-j} -1$.

In Lemma~\ref{lemma:well-formedConditions} we determine the conditions for the transformed state to be well-formed. To do so, we first formulate an equivalent but more intuitive requirement, which is condition (\ref{Lemma:conds1:SetC}) in the Lemma. 
It says that for each $j\ge 1$ there must be a value of $c$ such that $\hat{f}(c)=0$ with the $j$th term in \eqref{eqn:factorisedQFTOuputS} equal to zero and the first $j-1$ terms non-zero (in fact, by symmetry there must be $2^{n-j}$ such values). If for some $k$ there is no $c$ satisfying this condition, then there must not be any $c$ satisfying it for $j > k$ either, or the state will not be well-formed. 
In condition (\ref{Lemma:conds1:AmpConds}) we translate this notion into formal requirements about the relationships between the components of the untransformed input state components $\alpha_i,\beta_i$ which will ensure the transformed state will satisfy condition (\ref{Lemma:conds1:SetC}) and thus be well-formed.
Specifically, this requires each of the first $k$ input qubit to be (up to an arbitrary phase) in one of two superpositions which depend on the previous qubits, and excludes the remaining $n-k$ qubits from a similar set of possible states.

\begin{lemma}
	\label{lemma:well-formedConditions}
	Let $\ket{\psi_N}$ be a separable input state and $\ket{\hat{\psi}_N}=F_N\ket{\psi_N}$ be the transformed state. Then the following three conditions are equivalent:
	\begin{enumerate}[(i)]
		\item \label{Lemma:conds1:well-formed} $\ket{\hat{\psi}_N} \in \mathcal{V}_N$, i.e. the transformed state is well-formed. 
		\item \label{Lemma:conds1:SetC} There exists a $k\le n$ such that the set $$\mathcal{C}_j = \left\lbrace c \mid \forall{l \le j}\left( \alpha_l + e^{\pi i c/2^{l-1}}\beta_l = 0 \iff l=j \right)\right \rbrace$$ is non-empty for all $1 \le j \le k$ and empty for $k < j \le n$.
		\item \label{Lemma:conds1:AmpConds}
			 $(\exists{0 \le k \le n})(\exists {a_1\dots a_k \in \{0,1\}^k}) \left( \forall{1 \le j \le k}\left[\alpha_j = e^{\pi i \sum_{l=1}^j a_l/2^{j-l}}\beta_j \right]\right.$\\ 
			$\left.\qquad \qquad \qquad \qquad \wedge (\forall{a_{k+1}\dots a_n \in \{0,1\}^{n-k}})(\forall{n \ge j > k})\left[\alpha_j \neq e^{\pi i \sum_{l=1}^j a_l/2^{j-l}}\beta_j \right] \right).$
	\end{enumerate}

\end{lemma}
\begin{proof}
	
	(\ref{Lemma:conds1:well-formed}) $\implies$ (\ref{Lemma:conds1:SetC}): For any $x \in \mathcal{B}_N$, Definition~\ref{def:Bn} ensures that the number of ones in $x$, $\#_1(x) = 2^m$ for some $m\le n$, and hence the number of zeros, $\#_0(x) = 2^n - 2^m = \sum_{l=1}^{n-m}2^{n-l}$.
	 If $|\mathcal{C}_j| \neq 0$ then there exists a $c' \in \mathcal{C}_j$ such that $c'<2^j$ and $\hat{f}(c')=0$. But by symmetry we must also have $\hat{f}(m2^j+c')=0$ for $0\le m \le 2^{n-j} -1$ and hence $|\mathcal{C}_j|=2^{n-j}$.
	Also note that each $\mathcal{C}_j$ is disjoint by construction, and $\hat{f}(c)=0 \implies c\in \mathcal{C}_j$ for some $j$. 
	Hence, by construction, for a well-formed state we must have 
	$$\#_0\left(\mathcal{A}(\ket{\hat{\psi}_N} )\right)  \equiv \sum_{j=1}^n|\mathcal{C}_j| = \sum_{j:|\mathcal{C}_j| \neq 0}|C_j|.$$
	It follows that that for some $m$
	$$\sum_{l=1}^{n-m}2^{n-l} = \sum_{j:|\mathcal{C}_j| \neq 0}2^{n-j},$$
	which is satisfied if $\mathcal{C}_1 \dots \mathcal{C}_k$ are non-empty and $\mathcal{C}_{k+1}\dots \mathcal{C}_n$ are empty, with $k=n-m$.

	(\ref{Lemma:conds1:SetC}) $\implies$ (\ref{Lemma:conds1:well-formed}): In the first $K=2^k$ amplitudes, $2^{k-n}\sum_{j\le k}|\mathcal{C}_j| = \sum_{j=1}^k2^{k-j} = K - 1$ of them are zero. Let $\hat{f}(c')$ be the single one of these non-zero amplitudes. Then, by symmetry, $\hat{f}(dK + c')\neq 0$ for $0\le d \le 2^{n-k} - 1$. Thus, $\mathcal{A}( \ket{\hat{\psi}_N})=x^{2^{n-k}}$, where $x\in \{0,1\}^{K}$ and $\#_1(x)=1$. Any such $x$ is clearly well-formed, and thus the state $\ket{\hat{\psi}_N}$ is also well-formed.

	(\ref{Lemma:conds1:SetC}) $\iff$ (\ref{Lemma:conds1:AmpConds}): Note that $\sum_{l=1}^j a_l/2^{j-l}=\frac{1}{2^{j-1}}\sum_{l=1}^j a_l 2^{l-1}$, and we will proceed by induction for $j \le k$. 
	Since $\alpha_1 = e^{\pi i a_1}\beta_1 \iff \alpha_1 + e^{\pi i (1+a_1)}\beta_1 = 0$, such an $a_1\in \{0,1\}$ exists if and only if $|\mathcal{C}_1| \neq 0$.
	Now, assume that for all $1\le m < j \le k$, $\alpha_m = e^{\frac{\pi i}{2^{m-1}} \sum_{l=1}^m a_l2^{l-1}}\beta_m$ and $|\mathcal{C}_m| \neq 0 $.
	Then $$\alpha_j = e^{\frac{\pi i}{2^{j-1}}\sum_{l=1}^j a_l 2^{l-1}}\beta_j \iff \alpha_j + e^{\frac{\pi i}{2^{j-1}}(2^{j-1}+\sum_{l=1}^j a_l 2^{l-1})}\beta_j=0,$$ so such a bit string $a_1\dots a_j$ exists if and only if there is a $c$ such that $\alpha_j+e^{\pi i c/2^{j-1}}\beta_j=0$ (in fact $c=(2^{j-1}+\sum_{l=1}^j a_l 2^{l-1})\text{mod }2^j$). 
	Further, the inductive hypothesis ensures that for all $m < j$,
	\begin{align*}
		\alpha_m + e^{\pi ic/2^{m-1}}\beta_m &=\alpha_m + e^{\frac{\pi i }{2^{m-1}}(2^{j-1}+\sum_{l=1}^j a_l 2^{l-1})}\beta_m \\
		&= \alpha_m + e^{\pi i \frac{2^{j-1}}{2^{m-1}}}e^{\frac{\pi i}{2^{m-1}}( \sum_{l=1}^m a_l 2^{l-1})}\beta_m\\
		&= \alpha_m + e^{\frac{\pi i}{2^{m-1}}( \sum_{l=1}^m a_l 2^{l-1})}\beta_m\\
		&\neq \alpha_m - e^{\frac{\pi i}{2^{m-1}}( \sum_{l=1}^m a_l 2^{l-1})}\beta_m\\
		&= 0,
	\end{align*}
	thus such a bit string $a_1 \dots a_j$ exists if and only if $|\mathcal{C}_j|\neq 0$. 
	Hence, $\mathcal{C}_j$ is non-empty for $j \le k$ if and only if $\exists{a_1\dots a_k}\forall{1 \le j \le k}(\alpha_j = e^{\pi i \sum_{l=1}^j a_l/2^{j-l}}\beta_j )$.
	The condition that for $j>k$ and all $a_{k+1}\dots a_j\in\{0,1\}^{j-k}$ $\alpha_j \neq e^{\pi i \sum_{l=1}^j a_l/2^{j-l}}\beta_j$ is equivalent to $|\mathcal{C}_j|=0$, since $|\mathcal{C}_j|=0$ requires that there exists a $c$ such that $\alpha_j + e^{\pi i c/2^{j-1}}\beta_j =0$ and $\alpha_k + e^{\pi i c/2^{k-1}}\beta_k \neq 0$. The only $c< 2^k$ which satisfies this is $c=\sum_{l=1}^k a_l 2^{l-1}$, so by symmetry any $c$ which satisfies this must be able to be written as $c=\sum_{l=1}^j a_l 2^{l-1}$ for some $a_{k+1} \dots a_j$. Hence we see that (\ref{Lemma:conds1:SetC}) and (\ref{Lemma:conds1:AmpConds}) are equivalent. \qed
\end{proof}

Lemma~\ref{lemma:well-formedConditions} gives us conditions for when the first condition of Theorem~\ref{thm:genSepCond} is satisfied and it remains to determine which separable input states also satisfy the condition that $\mathcal{D}_{N'}(\ket{\hat{\psi}_N})$ is pair product invariant. The amplitudes which are deleted by the function $\mathcal{D}_{N'}$ are the $\sum_{l=1}^{k}2^{n-l}$ values of $c$ which are in $\mathcal{C}_j$ for some $j$.
In Lemma~\ref{lemma:PPIconditions} we work from Definition~\ref{def:PPI} to determine the conditions under which $\mathcal{D}_{N'}(\ket{\hat{\psi}_N})$ is pair product invariant.
\begin{lemma}
	\label{lemma:PPIconditions}
	Let $\ket{\psi_N}$ be a separable input state for which the transformed state $\ket{\hat{\psi}_N}$ is well-formed, i.e. $\ket{\psi_N}$ satisfies  the conditions of Lemma~\ref{lemma:well-formedConditions}. Let $k$ be as in Lemma~\ref{lemma:well-formedConditions} part (\ref{Lemma:conds1:AmpConds}), $n'=n-k$ and $N'=2^{n'}$. Then $\mathcal{D}_{N'}(\ket{\hat{\psi}_N})$ is pair product invariant if and only if for all $j>k+1$, $\alpha_j \beta_j = 0$, i.e. the $(k+1)$th qubit can be in an arbitrary superposition, and qubits $k+2$ to $n$ must not be in a superposition, although arbitrary phase is permitted.
\end{lemma}
\begin{proof}
	Let $c'$ be the smallest $c$ such that $\hat{f}(c) \neq 0$, and let $K=2^k$. By symmetry, the $N'$ non-zero amplitudes are $\hat{f}(dK + c')$ for $0 \le d \le N'-1$. The zero-deleted state is thus $\mathcal{D}_{N'}(\ket{\hat{\psi}_N})=(\hat{f}'(0),\dots,\hat{f}'(N'-1))$, where $\hat{f}'(d) = \hat{f}(dK + c')$. By breaking up the product in \eqref{eqn:factorisedQFTOuputS} we see that each of these amplitudes is of the form:
	\begin{align}
		\hat{f}'(d) &= \frac{1}{\sqrt{N}}\left[\vphantom{\prod_{l=1}^{n'}}\prod_{l=1}^k(\alpha_l + e^{\pi i (dK+c')/2^{l-1}}\beta_l)\right]\left[ \prod_{l=1}^{n'}(\alpha_{k+l} + e^{\pi i (d+c'/K)/2^{l-1}}\beta_{k+l}) \right]\notag\\
		&= \Gamma \prod_{l=1}^{n'}(\alpha_{k+l} + e^{2\pi i (d+\delta)/L}\beta_{k+l}),\label{eqn:zeroDelState}
	\end{align}
	where $L=2^l$, 
	$\delta = c'/K$ is independent of $d$, as also is $\Gamma = \frac{1}{\sqrt{N}}\prod_{l=1}^k(\alpha_l + e^{(2\pi i)^{dK/L}}e^{2\pi ic'/L}\beta_l)\neq 0$ (recall $k \ge l$ so $dK/L$ is a positive integer). 
	For all $j\in\{2,\dots,n' \}$, $m_1,m_2\in \{0,\dots, J/2-1 \}$, pair product invariance (recall Definition~\ref{def:PPI}) requires that both $\hat{f}'(m_1)\hat{f}'(J - m_1 - 1) = \hat{f}'(m_2)\hat{f}'(J-m_2-1)$ and $\hat{f}'(m_1)\hat{f}'(J/2-m_1-1)=\hat{f}'(m_2)\hat{f}'(J/2-m_2-1)$. Since each $\hat{f}'(d)\neq 0$, we require 
	\begin{equation}
		\label{eqn:PPIcondProof}
		\hat{f}'(J-m_2-1)\hat{f}'(J/2-m_1-1) = \hat{f}'(J-m_1-1)\hat{f}'(J/2-m_2-1).
	\end{equation}
	Symmetry means the left- and right-hand sides both contain common factors of $\Gamma^2$, as well as $j-1$ factors from the product \eqref{eqn:zeroDelState} for each transformed amplitude, due to the fact that $e^{2\pi iJ/L} = e^{\pi iJ/L}$ for $l<j$. Thus the condition \eqref{eqn:PPIcondProof} simplifies to 
	\begin{align}\label{eqn:PPIcondSimp}
		&\prod_{l=j}^{n'}(\alpha_{k+l} + e^{2\pi i (J-m_2-1+\delta)/L}\beta_{k+l})(\alpha_{k+l} + e^{2\pi i (J/2-m_1-1+\delta)/L}\beta_{k+l})\notag\\
		&=\prod_{l=j}^{n'}(\alpha_{k+l} + e^{2\pi i (J-m_1-1+\delta)/L}\beta_{k+l})(\alpha_{k+l} + e^{2\pi i (J/2-m_2-1+\delta)/L}\beta_{k+l}),
	\end{align}
	which holds for all $j,m_1,m_2$ if and only if $\mathcal{D}_{N'}(\ket{\hat{\psi}_N})$ is pair product invariant.
	
	We now show by induction that \eqref{eqn:PPIcondSimp} is satisfied if and only if for all $1 < j \le n'$, $\alpha_{k+j}\beta_{k+j}=0$. Firstly, consider the case that $j=n'$. The products in \eqref{eqn:PPIcondSimp} each contain only one factor, and expanding leaves only the cross-terms, and the condition simplifies to 
	\begin{align}
		&\alpha_{n}\beta_{n}(e^{2\pi i(N'-m_2)/N'}+e^{2\pi i(N'/2-m_1)/N'})
		 =\alpha_{n}\beta_{n}(e^{2\pi i(N'-m_1)/N'}+e^{2\pi i(N'/2-m_2)/N'}).\label{eqn:ppiBaseCase}
	\end{align}
	Since this must hold for all distinct $m_1, m_2$ only the trivial solution is possible, hence $\alpha_{n}\beta_{n}=0$.
	
	Now, assume that $\alpha_{k+l}\beta_{k+l}=0$ for $l=n',\dots, j+1$, $j>1$, and consider $\alpha_{k+j},\beta_{k+j}$. The products in \eqref{eqn:PPIcondSimp} run from $j$ to $n'$, but all factors for $l > j$ cancel when the pairs on each side are expanded since, by the inductive hypothesis, $\alpha_{k+l} \beta_{k+l}=0$ for these terms. The condition then reduces to a single factor and we find $\alpha_{k+j}\beta_{k+j}=0$ exactly as in \eqref{eqn:ppiBaseCase}. 
	
	Hence, the transformed state is pair product invariant if and only if for all $1 < j \le n'$ we have $\alpha_{k+j}\beta_{k+j}=0$.  \qed
\end{proof}

\begin{theorem}
	\label{thm:separableOutput}
	Given a separable input state $\ket{\psi_N}$, the transformed state $\ket{\hat{\psi}_N}$ is separable if and only if
	\begin{align*}
		(\exists{0\le k\le n})( \exists a_1\dots a_k &\in \{0,1\}^k ) \left( \forall{1\le j \le k}\left[\alpha_j = e^{\pi i \sum_{l=1}^j a_l/2^{j-l}}\beta_j \right]\right.\\ 
		&\left. \wedge \left(\alpha_{k+1} \neq \pm e^{\pi i \sum_{l=1}^{k} a_l/2^{k-l+1}}\beta_{k+1}\right) \wedge (\forall{n \ge j>k+1})\left[ \alpha_j \beta_j = 0 \right] \right)  .
	\end{align*}
\end{theorem}
\begin{proof}
	The proof follows directly from Lemmata \ref{lemma:well-formedConditions} and \ref{lemma:PPIconditions}.\qed
\end{proof}

Theorem~\ref{thm:separableOutput} allows us to determine if a given separable state $\ket{\psi_N}$ will be entangled or not by the QFT. While the set of such states which are not entangled by the QFT is infinite, the conditions are still highly restrictive, and there is only one qubit that can ever truly be in an arbitrary superposition. However, the conditions between each $\alpha_i$ and $\beta_i$ are relative, so separability of the transformed state is invariant under phase rotations of  individual qubits. These conditions, while restrictive, could be of value in developing new algorithms which make use of the QFT and give a strong insight into the entangling power of the QFT.

\subsection{Product-state De-quantisation}

For the set of states which are not entangled by the QFT, we can use the conditions of Theorem~\ref{thm:separableOutput} to extend the basis-state de-quantisation. Let $k$ be as in Theorem~\ref{thm:separableOutput}. Let $r=\sum_{j=2,\alpha_{k+j}=0}^{n-k}2^{-(k+j)}$ and $\omega = e^{2\pi i r}$ be the coefficient of $(\alpha_{k+2}+\beta_{k+2})\cdots(\alpha_{n}+\beta_{n})$ in $\hat{f}(1)$. The de-quantised algorithm for states which are not entangled by the QFT is the following ($\vec{b}[x]$ is the $x$th component of $\vec{b}$ starting from $0$):

\vspace{10pt}\noindent
{\bfseries Separable De-quantised QFT}\\
{\bfseries Input:} The $n$ two-component complex vectors $\vec{b_1}\vec{b_2}\dots \vec{b_{n}}.$\\
{\bfseries Output:} The $n$ transformed vectors $\vec{\hat{b}_1}\vec{\hat{b}_2}\dots \vec{\hat{b}_{n}}$.
\begin{enumerate}
	\item Calculate $k$, $a_1 \dots a_k$ as in Theorem~\ref{thm:separableOutput}
	\item Calculate $r$, $\omega$
	\item For $j=1$ to $k+1$:
	\item \quad Set $\vec{\hat{b}_{n-j+1}} = \oort \times \twoVec{\alpha_j + e^{\pi i \sum_{l=1}^{j-1} a_l/2^{j-l}}\beta_j}{\alpha_j - e^{\pi i \sum_{l=1}^{j-1} a_l/2^{j-l}}\beta_j}$
	\item End For
	\item For $j=1$ to $n-k-1$:
	\item \quad Let $l=n-j+1$
	\item \quad Set $\vec{\hat{b}_j}=\oort \times \twoVec{\alpha_l + \beta_l}{\alpha_l + \beta_l}$
	\item End For
	\item For $j=1$ to $n$:
	\item \quad Set $\vec{\hat{b}_{n-j+1}}[1] = \omega \vec{\hat{b}_{n-j+1}}[1]$
	\item \quad Set $\omega = \omega^2$
	\item End For
\end{enumerate}

\begin{theorem}
	\label{thm:prodAlgProof}
	The \emph{Separable De-quantised QFT} algorithm correctly computes the transformed $n$-qubit state $\ket{\hat{\psi}_N} = F_N\ket{\psi_N}$, where $\ket{\psi_N}$ is separable and the $c$th component of $\ket{\hat{\psi}_N}$ is described by \eqref{eqn:factorisedQFTOuputS}, and does so in $O(n)$ time.
\end{theorem}
\begin{proof}
	We first note that only one string $a_1\dots a_k$ can satisfy the first condition of Theorem~\ref{thm:separableOutput}: it is clear that only one value of $a_1$ satisfies it for $j=0$ and, for each subsequent $j\le k$, given $a_1 \dots a_{j-1}$ only one value of $a_j$ can satisfy the condition.
	The values of $k$ and $a_1\dots a_k$ can hence be found readily in $O(n)$ time by sequentially checking each pair $\alpha_j, \beta_j$ to see which option, $a_j=0,1$, makes the first condition of Theorem~\ref{thm:separableOutput} true, and setting $a_j$ accordingly. When neither is $a_j=0,1$ satisfies the condition we have found $k$.
	It is then evident that $r$ and $\omega$ can be efficiently found by direct calculation. 
	
	It remains to verify that the algorithm correctly produces the state 
	\begin{align*}
		\hat{f}(c)=&\frac{1}{\sqrt{N}}\prod_{j=1}^n(\alpha_j + e^{\pi i c/2^{j-1}}\beta_j)\\
		=& \frac{1}{\sqrt{N}}\left[\prod_{j=1}^{k+1}(\alpha_j + e^{\pi i c/2^{j-1}}\beta_j) \right] \left[\prod_{j=k+2}^n(\alpha_j + e^{\pi i c/2^{j-1}}\beta_j) \right].
	\end{align*}
	
	The algorithm calculates the amplitudes for each qubit, so if we let the $n$-bit binary expansion of $c$ be $c_n \dots c_1$ we have
	\begin{align}
		\hat{f}(c) &= \vec{\hat{b}_{1}}[c_n]\cdot \vec{\hat{b}_{2}}[c_{n-1}]\cdots \vec{\hat{b}_{n}}[c_1]\notag\\
		&= \frac{\omega^{c}}{\sqrt{N}}\left[\prod_{j=1}^{k+1} (\alpha_j + (-1)^{c_{j}} e^{\pi i \sum_{l=1}^{j-1}a_l/2^{j-l}} \beta_j) \right] \left[\prod_{j=k+2}^n (\alpha_j + \beta_j) \right]\notag\\
		&= \frac{\omega^{c}}{\sqrt{N}}\left[\prod_{j=1}^{k+1} (\alpha_j +  e^{\frac{\pi i}{2^{j-1}}(c_j 2^{j-1}+\sum_{l=1}^{j-1}a_l2^{l-1})} \beta_j) \right] \left[\prod_{j=k+2}^n (\alpha_j + \beta_j) \right]. \label{eqn:producedState}
	\end{align}
	Note that, since $\alpha_j=0$ or $\beta_j=0$, 
	$$\prod_{j=k+2}^n(\alpha_j + e^{\pi i c/2^{j-1}}\beta_j) = e^{2\pi i c r}\prod_{j=k+2}^n(\alpha_j + \beta_j) = \omega^c\prod_{j=k+2}^n(\alpha_j + \beta_j),$$
	so our algorithm produces this factor correctly.
	
	Since the output state is separable, the conditions of Theorem~\ref{thm:separableOutput} must be satisfied and only one out of the first $K$ amplitudes is non-zero. 
	This amplitude is the one with $c'=\sum_{l=1}^k a_l 2^{l-1}$, and by symmetry all the other non-zero amplitudes occur at $c=c'+ d2^{n-k}$ for $0\le d \le K -1$. To verify this, note that for all $j\le k$, we have $$\alpha_j + e^{\frac{\pi i}{2^{j-1}}\sum_{l=1}^k a_l2^{l-1}}\beta_j = \alpha_j + e^{\frac{\pi i}{2^{j-1}}\sum_{l=1}^j a_l2^{l-1}}\beta_j \neq 0,$$ and hence $\hat{f}(c') \neq 0$.
	From \eqref{eqn:producedState} it is clear that $\hat{f}(c)$ is calculated correctly for these values of $c$. For all other values of $c$ which have $c_1 \dots c_n \neq a_1 \dots a_n$, let $m$ be the smallest $i\le n$ such that $c_i \neq a_i$. Then we have 
	$$\alpha_m + e^{\frac{\pi i}{2^{m-1}}\sum_{l=1}^n c_l2^{l-1}}\beta_m = \alpha_m - e^{\frac{\pi i}{2^{m-1}}\sum_{l=1}^ma_l2^{l-1}}\beta_m = 0,$$
	and hence $\hat{f}(c)$ is correctly produced for all $c$.
	
 	The algorithm is also clearly seen to require $O(n)$ time, and thus the proof is completed.
	\qed
\end{proof}

This algorithm has all the advantages of the basis-state de-quantised algorithm, but operates on a much larger ranger of input states, making it a much more powerful de-quantisation. Importantly, just like the basis-state de-quantisation, it is actually more efficient than the QFT algorithm. While this algorithm will not work on all separable input states like the tensor-contraction simulation in~\cite{Yoran:2007aa}, it is a stronger de-quantisation in the sense that it gives a complete description of the output state as opposed to the probability of measuring a particular value, and is trivial to use as a subroutine in a larger de-quantisation.

\section{Discussion}\label{sec:discuss}

The ability to de-quantise the QFT algorithm brings up some interesting points. The two de-quantisations presented in this paper compute the Fourier transform on a restricted set of input states. On the other hand the standard QFT algorithm computes the Fourier transform on arbitrary separable or entangled input states. In fact, the standard QFT algorithm is a quantum implementation of the basis-state algorithm, but the linearity of quantum mechanics ensures that arbitrary input states are transformed by this simple algorithm. De-quantisation techniques such as the one presented, as well as those of \cite{Aharonov:2007aa,Browne:2007aa,Yoran:2007aa}, all have to efficiently simulate the linearity that is inherent in the quantum mechanical medium. The de-quantisations in this paper highlight the important distinction that should be made between the quantum Fourier transform and the quantum algorithm computing it. The QFT is a unitary transformation of an $n$-qubit state, while the QFT algorithm is a recipe for creating a sequence of local gates which computes the QFT on a given state. While these two notions are equivalent in quantum computation, when we depart from quantum mechanics this is no longer the case, and the de-quantised algorithm does not suffice to compute the complete QFT.

It is interesting to note that both de-quantisations presented in this paper run in $O(n)$ time, more efficient than the $O(n^2)$ of the quantum algorithm. This is due to the restrictions imposed by measurement no longer being present when we develop a classical counterpart. This increase in efficiency is something not seen in other de-quantisations of the QFT which are based on the quantum circuit topology, and thus inherently and perhaps unnecessarily work within the restrictions the quantum circuit was designed under. 
The Separable De-quantised QFT algorithm computes the QFT on a large number of input states, and any algorithm using a subset of these states can immediately be de-quantised using the algorithm presented.
The fact that both the input and output states are separable also ensures the existence of a de-quantised inverse algorithm too, which is of practical significance.
While it remains to be seen if any current algorithms can be de-quantised using the algorithm presented, any new algorithms developed will be able to be checked against the conditions to see if de-quantisation is possible.
Further, by looking for interesting algorithms working on states which remain separable, new classical algorithms might also be found.

Another issue worth noting is that we must be careful to consider the complexity involved in manipulating the complex amplitudes in a state-vector when performing de-quantisation. 
While our manipulation of complex amplitudes did not contribute to the complexity of the de-quantised algorithms presented in this paper, attention had to be paid to make sure this was the case.
If we had instead implemented directly the obvious algorithm and calculated each factor $\omega_j$ individually, this computation would have dominated the running time of the algorithm. 
In quantum computation, however, the amplitudes are just our representation of a property of physical states. 
It is these physical states, rather than the amplitudes, which are altered by unitary transformations, and as a result we observe the amplitudes changing. 
This reiterates the need for care when de-quantising, as the amplitudes have no a priori reason to be easily calculated, or computable at all for that matter.

\section{Summary}\label{sec:conclusion}
We have shown that the quantum algorithm computing the QFT can be de-quantised into a classical algorithm which is more efficient and in many senses simpler than the quantum algorithm, primarily because the need to avoid measurement of the system is no longer present. However, the direct de-quantisation of the QFT algorithm leads to a classical algorithm which only acts on a basis-state. This difference is due to the linearity of quantum computation ensuring a basis-state algorithm computes the complete QFT, highlighting this linearity as a key feature in the power of the QFT. By examining the entangling power of the QFT we derived conditions which ensure that the QFT leaves a separable state unentangled, and showed that this separability is invariant under phase-rotation of the input qubits. We extended our de-quantisation to work on this set of states without any loss of efficiency.

The restrictions on the amplitudes of the state vector for de-quantisation highlight symmetries in both the positions of zeros in the vector, and the relationship between non-zero amplitudes. These symmetries are invariant under the Fourier transform, and it is this invariance which makes de-quantisation possible.
This idea of looking for symmetries on separable states which are invariant is a promising technique for developing de-quantisations.

This de-quantisation of the QFT serves not only to illustrate more deeply the nature of the QFT, but also gives the possibility of de-quantising other algorithms which use it with very little effort.
Further, the results can help aid the creation of new quantum algorithms and subroutines by clarifying which symmetries lead to separability, and which do not; the latter offer the possibility of being exploited only by quantum algorithms.

\section*{Acknowledgements}
The author would like to thank Cristian S. Calude for many helpful discussions and much advice, Tania K. Roblot for comments and suggestions, and the anonymous referees whose comments helped improve the paper. This work was in part supported by a University of Auckland Summer 2010 Fellowship.

\bibliographystyle{model1-num-names}

\end{document}